\documentstyle[12pt]{article}
\begin{document}

\title
{\Large \bf 
Amplification of weak signals and stochastic resonance  via  on-off intermittency
with symmetry breaking
}

\author{ Changsong Zhou$^1$ and  C.-H. Lai$^{1,2}$ \\
        $^1$Department of Computational Science\\
        and $^2$Department of Physics\\
        National University of Singapore,
        Singapore 119260}

\date{}
\maketitle

\begin{center}
\begin{minipage}{14cm}

\centerline{\bf Abstract}
\bigskip

Nonlinear dynamical systems possessing reflection symmetry have an invariant
subspace in the phase space.  The dynamics within the invariant subspace 
can be random or chaotic. As a system parameter changes, the motion  transverse to  the
invariant subspace can lose stability, leading to on-off intermittency. 
Under certain conditions, the bursting behavior is symmetry-breaking. 
We demonstrate the possibility of observing multiplicative noise(chaos)-induced 
 amplification of weak  signal and stochastic resonance via on-off intermittency
with symmetry breaking in a  general class of symmetrical systems.   
Differences of this mechanism of stochastic resonance to that in noisy bistable
or threshold systems are discussed

PACS number(s): 05.40.-a, 05.45.-a

\end{minipage}
\end{center}

\newpage
\section{Introduction}

The phenomenon of stochastic resonance has been a subject of great interest since it was first proposed
in the study of the geophysical dynamics~\cite{bsv}. The idea
is that a signal can be amplified by the dynamics of a system in the presence of noise. 
The most frequently studied system is  the motion of a particle in a
symmetric double-well potential that is subjected to a periodic modulation and a Gaussian white
noise. The amplitude of the modulation is so small that, by itself, it cannot induce  any transition
between the two potential wells. On the other hand, the added  white noise can induce such 
transitions and controls the 
time-scale of the tunneling between the two potential wells. A resonance occurs when this 
time-scale matches the time period of the modulation. This phenomenon has been explored in various 
fields and many new applicability have been discoved, see Ref.~\cite{ghjm} 
for a  review of the phenomenon and further references. 
In general, three basic ingredients are required
for a system to display stochastic resonance, including a form of threshold or energy barrier, 
a weak coherent input and a
noise source  which is inherent in the system or added to the weak input. This source of ``noise''
 can be some form of  
chaotic dynamics in deterministic systems, because deterministic chaos resembles the features of noise on a
coarse-grained time-scale. With these features, the system can  display increased sensitivity to
the weak input at an optimal noise level.   

In this paper, we present a new  mechanism for realizing stochastic resonance in a general class of dynamical
system with reflection symmetry. Due to this symmetry, the system possess an invariant subspace in its phase
space. We are interested in the case where  the motion within this subspace is random or chaotic. As a
 system parameter changes, the stability of the subspace can be altered  and the system can display 
unusual dynamical behaviors, among which are on-off
intermittency~\cite{pst, phh, hph, ykl,cnts, yd} and
bubbling~\cite{abs,hcp,gb,vho}.   In on-off intermittency, the invariant
manifold is slightly unstable, and the system can remain close to
the invariant manifold for  long periods of  time,  interrupted only
 by some occasional large bursts away from the invariant manifold.
In bubbling the invariant manifold is stable; however, there are unstable
invariant sets  embedded in the chaotic sets. As a result, small perturbations
can result in large intermittent bursts from the invariant manifold.  
Another phenomenon that may accompany the onset of on-off intermittency
is symmetry-breaking,
in which the resulting  bursting behavior does not possess the system symmetry,
so that the system have two  coexisting symmetrical attractors.
 However, clear symmetry
breaking may not be observed in experiments because the system close
to the onset point  is very sensitive to external perturbations, such that  any 
small  noise in practice will induce transition
of the trajectories between the two symmetrical attractors, and symmetry is restored.

In this paper, we first study how small noise affects the transition
in  the system.
Then we investigate the response of the system to the input of  a stream of 
very weak binary signal, periodic
or aperiodic.
This weak input can also  induce transition of the trajectory between the two symmetrical components.
In the symmetry breaking region, the transition is determined by the applied weak input, and the 
output can be regarded as an amplification of the weak signal. The random or chaotic motion 
within the invariant subspace affects the response time of the system as well as the bursting 
frequency. As a parameter of this motion changes, we can observe the phenomenon of stochastic
resonance. 
Since the source of noise from the random or chaotic motion is multiplicative to the motion
transverse to the invariant subspace, we call the resonant phenomenon  multiplicative 
noise(chaos)-induced stochastic resonance. We employ various  measures to characterize 
this phenomenon in
the context of different possible applications, including residence-time distribution, bit error
probability and  amplification factor.

\section{ The System} 

We consider the following general class of systems:
\begin{eqnarray}
x_{n+1}&=&f(x_n),\\
y_{n+1}&=&F(x_n,p)G(y_n),
\end{eqnarray}
where  $f(x)$ is a noise generator or a  chaotic process.  
The function $G(y)$ possesses the  
symmetry $G(-y)=-G(y)$, thus the subspace ${y=0}$ is invariant. 
$F(x_n,p)$ is a certain coupling function  which can be regarded as a multiplicative driving 
to the subsystem $y$, and  $p$ is a tunable parameter of the driving. 
 The stability of the invariant subspace $y=0$ is determined by the transverse
Lyapunov exponent
\begin{equation}
\lambda=\lim\limits_{N\to \infty}\frac{1}{N}\sum\limits_{n=1}^{N} 
\ln |F(x_n,p)G^{\prime}(0)|=\langle \ln |F(x, p)|\rangle,  
\end{equation}
where $G^{\prime}(0)=dG(y)/dy|_{y=0}$ is a constant. By absorbing this constant into the 
function $F(x_n, p)$, one can always set  $G^{\prime}(0)=1$. 
Let $D_x=\langle (\ln |F(x, p)|-\lambda)^2 \rangle$ denotes the  
variance of $\ln |F(x, p)|$.  
The stability of the invariant subspace is determined by the parameter $p$, and  $\lambda=0$ 
defines the critical value  $p_c$ which is the onset  point of on-off intermittency.

\subsection{Brownian motion model for on-off intermittency and bubbling}

To understand the unusual behavior of the system close to the critical point of the stability,
let us examine the linear dynamics of $y$ close to the invariant solution $y=0$, e.g. $|y|\le \tau$,
namely, 
\begin{equation}
y_{n+1}=F(x_n,p)y_n.
\end{equation} 
A state $|y|\le \tau $ is referred to as a laminar phase and $|y|>\tau$ a bursting phase. Here $\tau$ is a
small enough value so that the linear approximation in Eq. (4) is valid.
Introducing the variable $z=\ln |y|$, we get
\begin{equation}
z_{n+1}=z_n+\lambda+\sqrt{D_x}\xi_n,
\end{equation}
where $\xi_n=(\ln |F(x_n, p)|-\lambda)/\sqrt{D_x}$ is a random or chaotic variable with a mean 0 and a variance 1. 
Now if we rescale the system to a coarse-grained time-scale  by a factor of $N$  as
\begin{equation}
z^{\prime}_n=\frac{z_{nN}}{N}, \;\;\;\; \xi^{\prime}_n=\frac{1}{N}\sum_{i=0}^{N-1} \xi_{nN+i},
(n=0,1,2, \cdots),
\end{equation}
we can see that 
\begin{equation}
z^{\prime}_{n+1}=z^{\prime}_n+\lambda+\sqrt{D_x}\xi^{\prime}_n,
\end{equation}
which has the same dynamics as Eq. (5).  If the random or chaotic signal $x_n$ has
very short correlation time,  $\xi^{\prime}_n$ will have an asymptotic Gaussian distribution for large $N$ 
according to the central limit theorem. In this context, the driving signal $x_n$ can be viewed as a kind of
multiplicative noise to the motion of $y_n$.

Based on the above consideration,  
to analyze the long time behavior, map (5) can be replaced by the
corresponding stochastic differential equation
\begin{equation}
\frac{dz}{dt}=\lambda+\sqrt{D_x}\xi,
\end{equation}
where $\xi$ is a Gaussian white noise with a normal distribution $N(0,1)$.
This equation describes one-dimensional  Brownian motion  with a constant drift $\lambda$ and a diffusion
coefficient $D_x/2$.    
When $\lambda$ is slightly positive, the motion drifts to the positive direction,
leading to repulsion from the invariant subspace, but the 
 diffusion may make  the motion  access deeply into the negative values of $z$, so
that the system  can come in   and remain close to the invariant subspace for some long period of time,
leading to  on-off intermittency. For $\lambda$ slightly negative, the motion will eventually
drift to $z\to -\infty$;
however, if there is small noise added to $y_n$, 
the motion 
is prevented from drifting to $z\to -\infty$. The effects of the additive noise to    
Eq. (2) can be modeled by a reflecting boundary condition~\cite{phh, cl} of 
the Brownian motion in Eq. (8).
 With  this reflecting
boundary, $z$ may  access  positive values due to the diffusion, leading to  
the behavior of attractor bubbling.
In the presence of perturbations, on-off intermittency and bubbling are essentially the same phenomenon, and
we shall refer to  on-off intermittency  from now  on. 
When focusing on the laminar
phases, the nonlinearity of the function $G$ is not important. It only serves to keep the state $y$ 
bounded.    

\subsection{Symmetry breaking and conserving}

The phenomenon of symmetry-breaking, however, is associated with the nonlinearity of the function $G$.
In Fig. 1, we show two possible situations of the system  behavior.    
Let $Y_1>0$ and $Y_2>0$ the values of $y$ at which $G(Y_1)=\hbox{max}[|G(y)|]$ and $G(Y_2)=0$. Plot (a) 
depicts the situation where  $y_{max}=\hbox{max}[|F(x,p)|]G(Y_1)<Y_2$, so that a trajectory starting with
$0<y_0<Y_2\;(-Y_2<y_0<0)$ will remain in the positive (negative) part forever
 in the noise-free case, and 
the system is symmetry-breaking since the trajectory does not possess the reflection symmetry of the function 
$G$. While in plot (b), $y_{max}>Y_2$,  and the system has only one asymptotic 
attractor possessing the reflection symmetry, thus is symmetry-conserving. 
A transition from one symmetrical component to the
other  happens whenever $|y_n|>Y_2$.  The point where  $y_{max}=Y_2$  is a 
symmetry increasing bifurcation point\cite{l}.  

To study the property of the  symmetrical dynamical system subject to small
noise or coherent  signal, we employ a function 
$G$ in which $Y_2$ is adjustable independently. Since many properties observed in the following  are      
quite common in this class of systems, we choose in this paper a piecewise linear function for
simplicity of analysis. 
\begin{equation}
G(y)=\left\{
\begin{array}{ll}
\frac{1}{c}(-1-c-y),& y<-1,\\
y,                         & |y|\leq 1,\\
\frac{1}{c}(1+c-y),& y>1. \\
\end{array}\right.
\end{equation}
Also since the properties of on-off intermittency are universal for many different driving signals $F(x_n, p)$,
we simply  let $F(x_n, p)=px_n$, and  use uniform random numbers $x_n\in (0,1)$ in simulations. 
With this implementation of the system, 
one has $\lambda=\ln p-1$,  $D_x=1$, $y_{max}=p$ and  $Y_2=1+c$ so that the symmetry 
increasing bifurcation occurs at $c_s=p-1$. The onset point of 
on-off intermittency is $p_c=e=2.71828...$.  

In the following several simulations,  we take 
$p=2.8$ above the critical point if not explicitly pointed out. 
Fig. 2 shows typical behavior of the system with symmetry breaking ($c=2$) and
symmetry conserving ($c=1$).

\subsection{Sensitivity to small perturbations}

Before we demonstrate the sensitivity of the system to weak  coherent signal and the phenomenon of stochastic
resonance,  we show how the system behaves in the presence of additive small noise.  
Small noise sets a reflecting boundary to the Brownian motion and can change the property of the laminar
phases considerably~\cite{phh, cl}. 
Noise has another effect on the system 
behavior: it can induce in the trajectory transitions between the two symmetrical 
components of the system, and the symmetry breaking will not manifest. As an example,
Fig. 3  shows  system  behavior corresponding to Fig. 2, but  with an additive
small noise 
\begin{equation}
y_{n+1}=px_nG(y_n)+e_n.
\end{equation}    
The standard deviation of the Gaussian white noise $e_n$ is $\delta=10^{-5}$. 
In the presence of even  very small noise, symmetry is restored for the originally 
symmetry-breaking system. Since in practice noise is inevitable, clear symmetry breaking 
as in Fig. 2(a) cannot be observed in real experiments. 

The bursting patterns in Fig. 3(a) and (b)
are qualitatively different. In Fig. 3(a), transition between $y>0$ and $y<0$ occurs only when
$y$ come to the noise level, while in Fig. 3(b) transition can occur both in  laminar phases  and 
bursting phases.

\section{Amplification of  weak signal and stochastic resonance}

In most previous studies on stochastic resonance, the noise induces hopping  between the states separated by 
a  barrier or a  threshold with an average
waiting time $\langle T_r\rangle$. When  $\langle T_r\rangle$ is comparable with half the period of the
applied weak signal, this noise-induced hopping becomes statistically synchronized with the weak signal, and
stochastic resonance ensues.  In our system, the mechanism is different.  We do not need  additive noise
to generate transitions. A weak signal by itself  can generate such transitions. 

Let us first consider the response of the system to a weak periodic binary signal:
\begin{eqnarray}
y_{n+1}&=&px_nG(y_n)+s_n,\\
s_{n+2T}&=&s_n=\left\{ \begin{array}{rl}
A,& 0<n\leq T,\\
-A,& T<n\leq 2T.
\end{array}
\right.
\end{eqnarray}
In the example  in Fig. 4, we can see astonishing difference of the system 
response to a weak  periodic signal with amplitude
$A=10^{-5}$ and bit duration $T=1000$. It is clear that for the symmetry-breaking case, the transition
is now almost totally  governed  by the  switching of the periodic  binary signal. 
A weak sinusoidal input will produce similar output. 
While for the symmetry-conserving 
system,  there are many additional transitions induced by bursting phases in the duration of a signal bit,
and  the output  do not have the clear periodicity of the signal.

To characterize
the difference in the bursting pattern,  we calculate the distribution of the residence time.
The residence time  $T_r$ is defined  as follows: starting with  a moment $n_0$ when the system
produces for the first time a large bursting state, say $y_{n_0}<-y_{th}$, $n_1$
is the subsequent time when
the system first produce a bursting state $y_{n_1}>y_{th}$ and $n_2$ is the time
when $y_{n_2}<-y_{th}$ again,
and so on; the quantity $T_r(i)=n_i-n_{i-1}$
represents the residence time the system stays in one of
the symmetrical component between two subsequent transition events.
$y_{th}$ is a value of the order of $y_{max}$.  In all our simulations in this paper, we set
$y_{th}=1.0$.

Fig. (5) shows the  residence time  
distribution $P(T_r)$ of the bursting behavior  in Fig. (3). For the
symmetry-breaking system, $T_r$ distributes around the bit duration $T$     
of the input signal. While for the symmetry conserving system, the distribution is almost the same as
that of the input of noise (not shown). 

We should point out that for symmetry-breaking system, a single peak of $P(T_r)$ around $T$ occurs 
only for $T$ large enough. For $T$ smaller than a certain value, the distribution $P(T_r)$ begins to show other
peaks  centered at odd multiples of the bit duration $T$, i.e. at $(2k-1)T\; (k=1,2...)$, 
and  the smaller the $T$, the greater the number of  peaks. The peak height decreases
exponentially with  $k$. This property is quite common in stochastic resonance systems~\cite{ghjm}. 
Fig. 6  shows the results of $P(T_r)$ for $T=1000$ and $T=200$. 
For $T=1000$, a very small peak at  $k=2$ begins to show up, while for $T=200$, seven peaks are clearly
discernible.  This behavior is associated with the relaxation
time $T_0$ of the system, which is the time for the system to relax to the  statistically stationary 
state after the
signal switches from one value to the other. When $T_0>T$, the transition of the system is not able to 
keep up with  the switching of the signal. The appearance of the peaks can be simply explained: the system may perform a
transition   after  
the signal switches from a bit to the other, thus $T$ is a preferred residence time. If 
the system fails to perform a transition before the
signal switches back, it has to wait for a full period before it can produce a transition, and the second
peak is therefore located at $3T$, and so on.  

The behavior of the system with a weak signal  is determined by the competition between the strength of
 the  constant drift $\lambda$ and the
diffusion $D_x/2$.
If $p$ is far
 away below the critical point,  the system will quickly relax to a signal-induced metastability state,
 and can  rarely produce large bursting states, 
and the bursting  pattern 
may not possess the periodicity of the weak signal. With $p$ approaching $p_c$, the system can 
produce  large  bursting states more frequently, and  the periodicity of the signal can be manifested.
 When $p$ is far away above  the critical point,  
the  system can seldom have  access to the level of weak input $|y|\sim A$, 
and the switching 
of the signal can no longer determine the transition of $y$; the  bursting pattern loses the periodicity of the
signal again. An optimal response will be obtained around the critical point where the drift is
dominated by the diffusion,  and  
the  system can have  access to the level of the weak input so that the transition is sensitive to 
the switching  of the weak signal, and at the same time can produce large bursting states quite quickly.

In order to quantify the  the response of the system as a function of $p$, we introduce the area under the
 peak  centered at the bit duration $T$ in the residence time distribution as a 
measure of the strength of the periodicity:
\begin{equation}
S=\sum\limits_{T-T/4}^{T+T/4}P(T_r).
\end{equation}
$S$ as a function of $p$
is shown in Fig. 7.  It increases with $p$, reaches a maximum and decreases again, displaying the 
typical feature of
stochastic resonance. This feature is quite robust to noise additive  to the weak periodic signal,
 as seen in Fig. 7 where $S$ as a function of $p$ is also shown for different level of additive noise.   
When $p$ increases, the system may  enter into the symmetry-conserving region, and the transition during  the
bursting phases will degrade the periodicity of the output. Now if $c$  has  such a  
value that the system enters into the symmetry-conserving region before reaching the resonant point, the
maximal value of $S$ will  be  slightly after  the symmetry increasing point $p_s=1+c$,
because the transition induced by the bursting phases is rare just beyond this point, and it become more
significant for $p$  going deeper into the symmetry converging region. An example of this case is shown
in  Fig. 7 for $c=1.5$, where the maximal $S$ is found around $p=2.55$. This optimal response is not
a resonant behavior in the sense discussed above. However, in a wider sense, it can also be 
regarded as a resonant phenomenon because it  results  from the competition between the  transition induced
by the weak signal and that induced by the bursting phases.

In the above, we characterize the stochastic resonance by the  residence time distribution  for a periodic
binary signal. One should have noticed that the  periodicity of the
signal is not important for observing the resonance phenomenon. However, if we consider aperiodic signals, for example
a random series of binary bits, we should employ some other quantities to quantify the phenomenon. 

First, let us consider a possible application of the system in the detection of a weak signal with bit duration
$T$. The detection is performed as follows: we look at the large bursting state $|y|>y_{th}$; if the sum of
these states in a bit duration is positive (negative), then this  bit is detected as $+1 \;(-1)$; if this
number is zero, we cannot make a decision. With this scheme, one may detect a very weak signal with 
a low-resolution detector. We calculate the probability of bit error $P_e$ as a function of $p$. The result is shown in 
Fig. 8. A random stream of signal with $10^6$ bits is used in simulation. When $p$ is quite below the
critical point, the system produces large bursts quite sporadically, and only a very small portion of the bits is 
detected. The bit error probability is  close to $1.0$. On the opposite hand, the system  produces large 
bursting states quickly, but is not sensitive to the signal, 
and $P_e$  tends to 0.5. An optimal  detection with smallest bit error probability 
is obtained around the critical point. We also calculate $P_e$ in the presence of additive noise.  
Again we see that the system 
is very robust to additive noise. In this sense, the system may find  application in the detection of 
a weak signal buried in relatively high level of noise.  
    
Next let us consider the application of the system as an  amplifier of the weak signal. 
A natural measure of the output
is the ensemble average $\langle y_n \rangle$. An amplification factor $I$ can be defined 
as
\begin{equation}
I^2=\lim_{N\to \infty}\frac{1}{N}\sum_{n=1}^N \frac{\langle y_n \rangle^2}{A^2}.
\end{equation}
For a periodic signal, 
\begin{equation}
I^2=\frac{1}{T}\sum_{n=1}^T \frac{\langle y_n \rangle^2}{A^2}.
\end{equation}

For this system, we can give an analytical estimation of the amplification factor under the adiabatic
condition $T\gg T_0$. To perform this analysis, 
let us recall the Brownian motion model
in Eq. (8). The probability distribution of the variable $z$ satisfies the Fokker-Planck equation 
\begin{equation}
\frac{\partial W}{\partial t}=-\lambda \frac{\partial W}{\partial z}+\frac{D_x}{2}
\frac{\partial ^2 W}{\partial z^2}.
\end{equation}
The small signal and confinement of the nonlinearity can be modeled by two reflecting
boundaries of the Brownian motion.  If $T\gg T_0$,  
the probability distribution of $z$ will establish a stationary state during a bit duration,
namely, $W(z)=C\exp(\alpha z)$,
where $\alpha=2\lambda/D_x$.
In the original variable $y$, it becomes
$W(y)=C|y|^{\alpha-1}$.
If the system is symmetry breaking, after reaching the stationary state, $y$ is always positive (negative)
for signal bit $+1\;(-1)$.  
With the normalization  condition  
\begin{equation}
\int_A^{y_{eff}}Cy^{\alpha-1}dy=1,
\end{equation}
where $y_{eff}$ is a parameter used to  represent the  reflecting boundary due to   
the  nonlinearity of the system,  
we can estimate the amplitude of the  ensemble average 
$\langle y_n \rangle $ of  the system with a weak  input (bit $+1$)  as 
\begin{equation}
\langle y \rangle\approx \frac{\int_{A}^{y_{eff}}yW(y)dy}{\int_{A}^{y_{eff}}W(y)dy}
=\frac{\alpha}{1+\alpha}\frac{y_{eff} \beta-A}{\beta-1},
\end{equation}
where $\beta=(y_{eff}/A)^{\alpha}$.  From Eq. (18),   $\langle y \rangle$  is a monotonic 
increasing function of $p$ and the stochastic resonance cannot be observed  should the  adiabatic
condition hold true for any value of $p$.  

Close to the  critical point $p_c$, $|\alpha|\ll 1$, the diffusion is dominant and the system can produce large 
bursting states quickly. As an estimation, we can just take
$y_{eff}=y_{max}$.  If $|\alpha| \ln(y_{max}/A)\ll 1$, one has $\beta\approx 1+\alpha \ln(y_{max}/A)$. For weak 
input  $A\ll y_{max}$, we arrive at 
\begin{equation}
\langle y \rangle\approx \frac{y_{max}}{\ln (y_{max}/A)}, 
\end{equation}
which decreases to zero  with the decrease of the signal amplitude $A$ only logarithmically.    
The amplification factor in this case is just  
\begin{equation}
I=\frac{\langle y \rangle}{A}=\frac{y_{max}}{A\ln (y_{max}/A)}, 
\end{equation}
which shows that the system is very sensitive to weak signal close to the critical point.  
This feature of sensitivity is quite different from  that of the sensitivity near the onset
of a period-doubling bifurcation  in many dynamical systems~\cite{wm}.
There the system is only sensitive
to perturbations near half the fundamental frequency of the system for bifurcation parameter
very close to the onset point.

This model analysis is  demonstrated by simulation. 
Fig. 9 shows $I$ as a function of $m$  for small signal amplitude $A=10^{-m}$. 
The parameter $p=2.72$ in this  simulation
is very close to the critical point and $T\gg T_0$ is satisfied. 
It is seen that the  analytical estimation of Eq. (20) with $y_{max}=p$ fits  the  simulation result very 
well  for $A$ covering  several orders. 

If the adiabatic  condition is not satisfied, the system may not establish  a stationary state during a bit 
duration, and the transient behavior during the relaxation process becomes  significant. 
The time-dependent probability distribution is 
not easy to obtain, and we rely on simulations to estimate $I$. Although the measure $I$ is
applicable to aperiodic signals,  we employ periodic signal in our simulation, because  
it can reduce considerably the computation effort.

Now let us investigate how $I$ changes with $p$. 
As stated in the above, 
if $p$ is far away below the critical point,   
the constant drift $\lambda=\ln p-\ln p_c<0$  makes the system stay most of time close to the low 
boundary $|y|\sim A$.  
The amplitude of the ensemble average  is quite small. 
It  increases quickly when $p$ approaches the critical point where the
diffusion become dominant. 
However, $T_0$ is also increasing,  and $\langle y_n\rangle$ may not reach the maximal value in Eq. (18) 
when the system is not quick enough to follow the modulation of the signal, and $I$ may begin to 
decrease. There will be an optimal amplification of the signal. Since the relaxation time is longer
for smaller $A$, the stochastic resonance peak  shifts to smaller value of $p$ for smaller $A$.  
 The amplification will be degraded 
further if the system enters into the symmetry conserving region.    

Typical behavior of the system   with respect to the parameter $p$ is illustrated in Fig. 10.
In this simulation,
we take $c=2.2$ so that the system will move into the symmetry-conserving region at
$p_s=1+c=3.2$ which is high  above the critical point.
Fig. 10(a) shows time series of $\langle y_n \rangle$ for several typical values of $p$.
The dependence of $I$ on the parameter
$p$ is shown  in Fig. 10(b) for different values of $c$ and signal amplitude $A$. Complying  with  the above
analysis, $I$ increases with $p$ untill $T_0$ becomes long enough or untill the system moves into 
the symmetry-conserving region.  We will observe the optimal system response to the small signal 
in the region $p<p_s$  if $T_0$  becomes significantly longer than $T$  before
the system enters into the symmetry-conserving region, as for the case $A=10^{-7}$ in Fig. 10(b) where the
maximal $I$ is found around $p=2.95$ before $p_s=3.2$. 
The maximum will be slightly after the point 
$p_s$ if the  system  enters into the symmetry-conserving region before  reaching the resonant point, 
as for the case of  $A=10^{-5}$ and $c=1.8$ in Fig. 10(b).

Again we examined  the robustness of the  property of stochastic resonance to additive noise.
In Fig. 11, $I$ as a function of $p$ for different level of additive noise is shown.
Amplification of the small signal is obtained even if the noise level is much higher than the signal 
amplitude.  

Since  amplification of weak  signal and stochastic resonance in the system can be understood  
by the Brownian motion model derived from the linear dynamics close to the invariant subspace, 
this phenomenon is universal in a general class of symmetrical systems with random or chaotic
motion within the subspace.

\section {Discussion}

We demonstrate  a new mechanism of  stochastic resonance in a general class
of symmetrical dynamical systems with on-off intermittency and symmetry breaking. 
The system has an invariant subspace whose stability is determined by the random or chaotic motion
within the subspace. Close to the critical point of the stability, the system is very sensitive to 
small perturbations, since the  state $y$ has a  power-law distribution  in a wide interval
$10^{-m}<|y|<y_{eff}$. The behavior of the system response to a weak binary signal can be understood
by the competition between the drift and the diffusion in the Brownian motion model derived from the linear
dynamics closed to the invariant subspace.  
When the diffusion becomes  dominant  for the parameter  $p$ close to the critical point of stability, 
the system attains 
the ability of amplification of a very weak signal, not via  additive noise, but via multiplicative noise.
When the parameter $p$ determining  the stability varies across the critical point from  below to 
above, the sensitivity to the weak signal increases and then decreases after reaching a maximum, displaying
the phenomenon of stochastic resonance. The resonance occurs when the system can have access to the
level of the weak signal to become sensitive to the switching of the signal,  and produce large output 
quickly at the same time. 

There  are several differences  of this mechanism of stochastic resonance  
to  that in a noisy bistable or threshold  system.

(1) The source of noise is not additive but multiplicative, which is inherent from the random or chaotic
 motion within the invariant subspace. 

(2) In a noisy bistable or threshold  system, the weak signal by itself cannot induce  hopping over the
the energy barrier or the threshold.  The additive noise induces such tunneling process which introduces
a noise-controlled  time-scale into the system. When this time-scale matches that of the weak signal,
the statistical synchronization of the tunneling to the signal leads to resonant  
response to the signal in the system. 
In the present
mechanism of stochastic resonance, the multiplicative noise does not  induce transition between the two symmetrical
component,   and there  does  not exist a noise-induced  time scale of the transition. The transition is induced
by the weak signal itself in the symmetry breaking case.  A change of the level of the multiplicative noise
changes the response time of the system. The resonance occurs not because this response time matches the period  of
the signal, but because the system responds quickly enough to follow the switching  of the signal and produces large
output quickly  after a switching .  The resonant phenomenon is not restricted to periodic signals. 
We have employed different
quantities rather than the conventional signal-to-noise ratio suitable for a  sinusoidal input to quantify the
phenomenon.

(3) In our systems  with symmetry breaking, although  there exist two distinguished symmetrical parts  in the
system, there is not any clear form of threshold or barrier in the system.

The phenomenon demonstrated in this paper is  universal in a class of systems. Many systems, for example 
symmetrically coupled identical chaotic systems~\cite{vho,yf,pc,j} or interacting stranger attractors~\cite{pik}, 
can be reduced to a form  similar to 
the systems discussed  in this paper, and we may expect to observe similar phenomenon around the critical point of
synchronization. However, in such systems, a change of the parameter of the coupling strength may 
change the  chaotic dynamics within the synchronization manifold, and this can 
bring about additional complication into the system response to a weak signal. This will be a topic for future
study.    

The behavior of on-off intermittency or bubbling may be harmful in some applications, such as secure
communication using synchronization of chaos~\cite{co}, because in practice high-quality synchronization can be 
 destroyed by large intermittent bursts from the synchronization manifold due to unavoidable 
perturbations~\cite{abs, hcp, gb, vho}.
We demonstrate that such behavior, however, can be employed to amplify extremely weak signal, and the
amplification is shown to be very robust to additive noise. This may lead to useful applications of the
behavior which is quite universal in many systems.

\bigskip
{\bf Acknowledgements:}

This work was supported in part by research grant No. RP960689 at the National
University of Singapore.  Zhou is supported by NSTB.

\newpage
{\large \bf Figure Captions}
\begin{description}

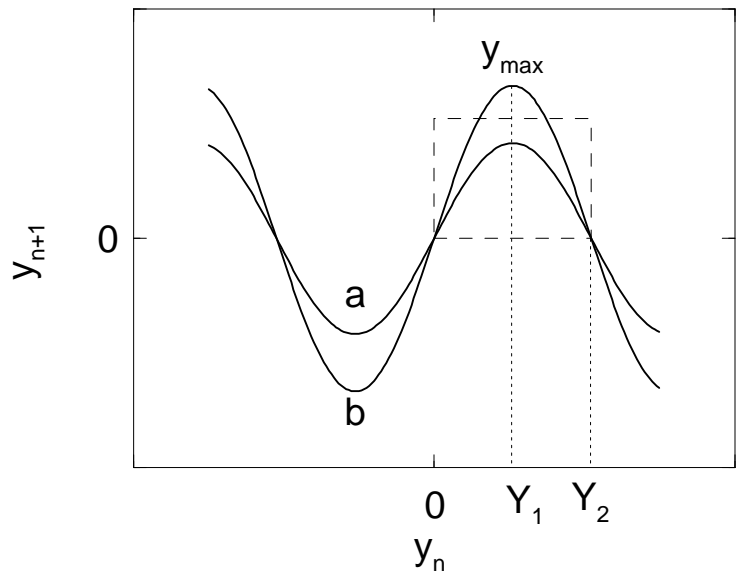
\item Fig. 1. Illustration of  two cases in the system: (a)  $y_{max}<Y_2$, symmetry breaking and (b)
$y_{max}<Y_2$, symmetry conserving. 

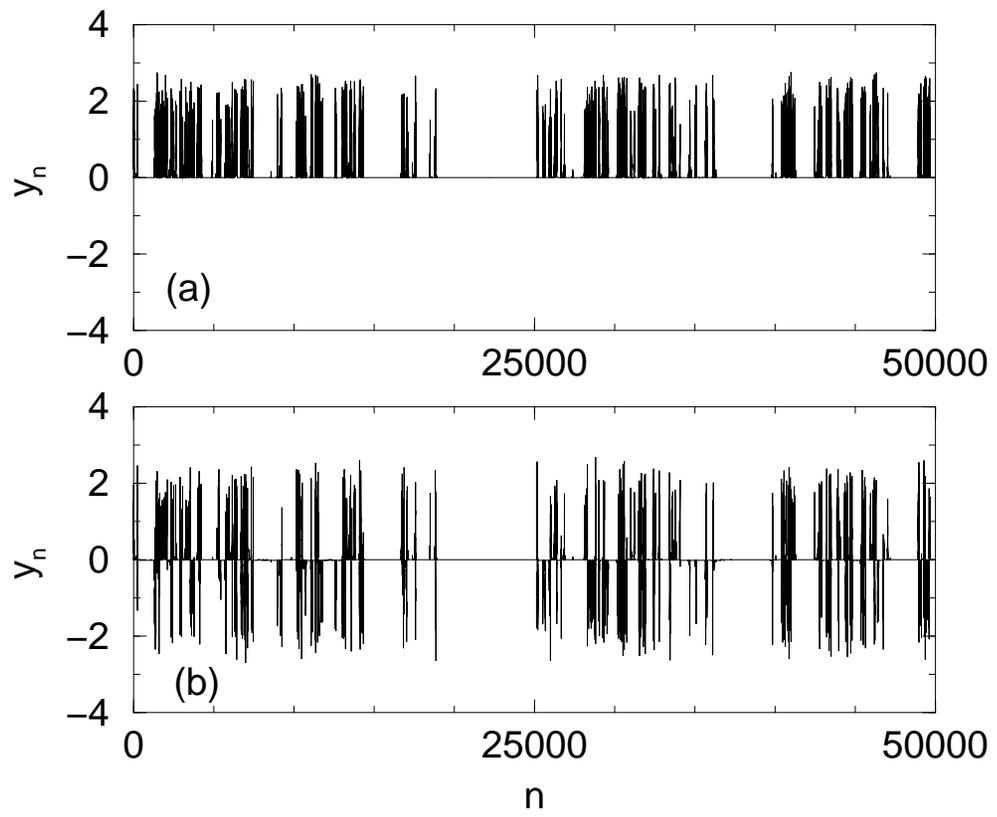
\item Fig. 2. Typical time series of on-off intermittency. (a) $c=2$, symmetry  breaking and (b)
$c=1$, symmetry conserving. 

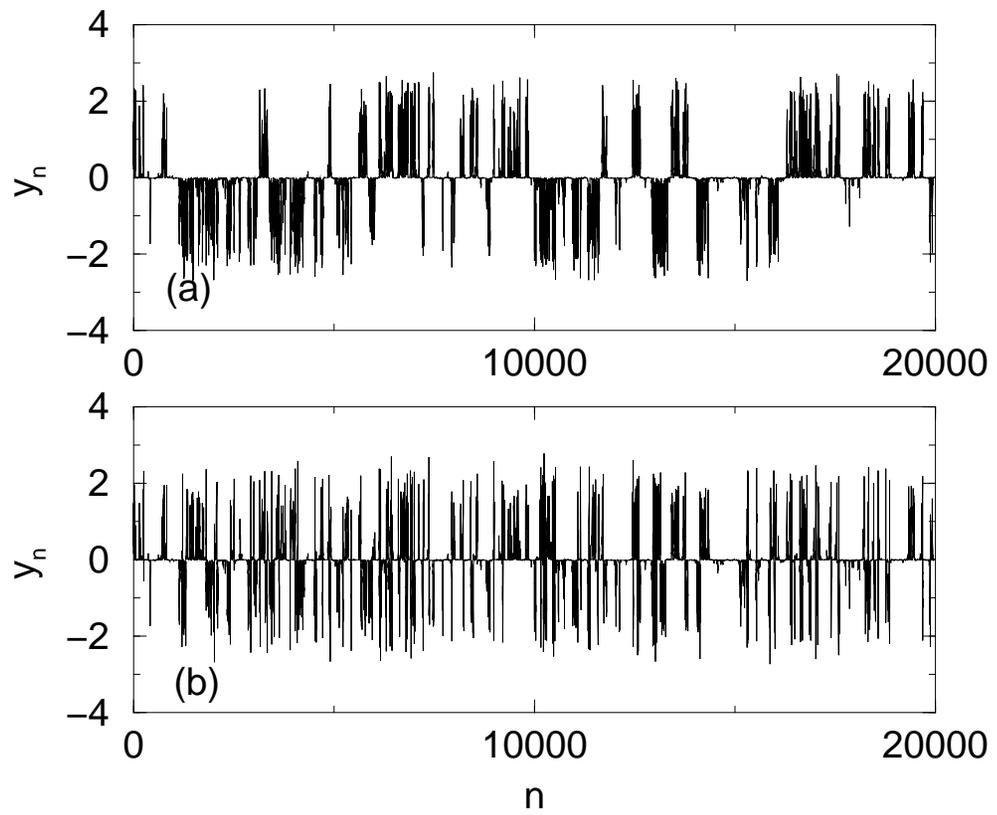
\item Fig. 3. Typical time series of on-off intermittency in the presence of small noise with
$\delta=10^{-5}$. (a) $c=2$, symmetry  breaking and (b) $c=1$, symmetry conserving.

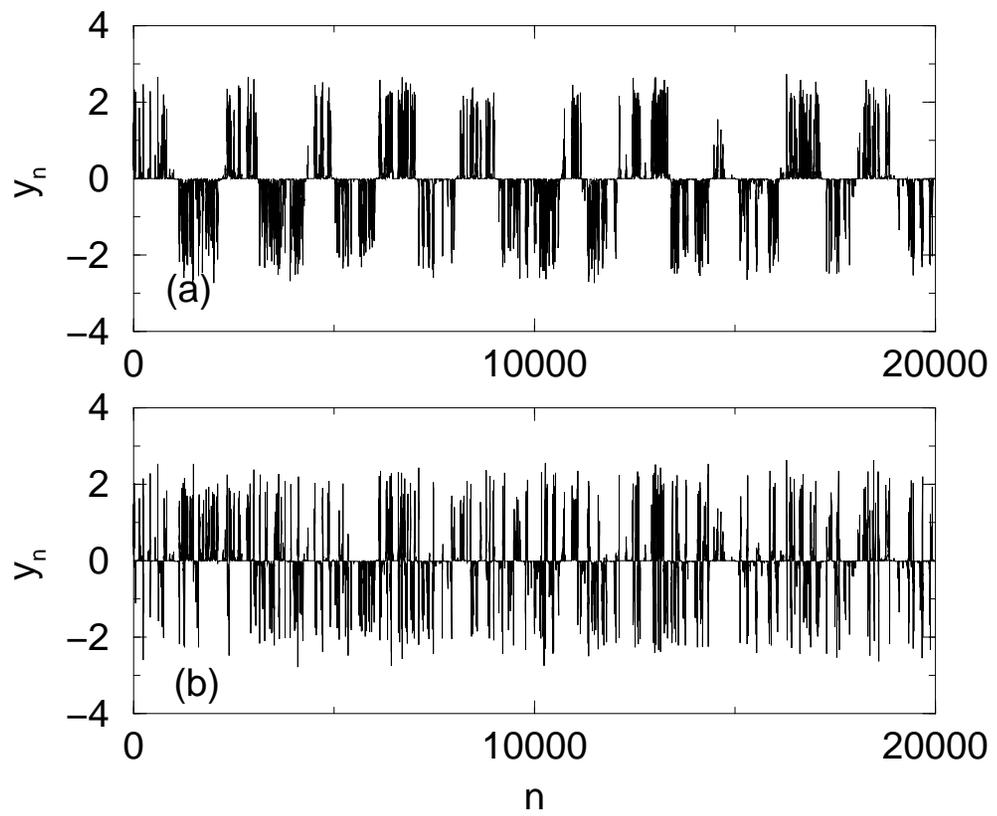
\item Fig. 4. Typical time series of on-off intermittency in the presence of weak  periodic signal.
with  amplitude $A=10^{-5}$ and bit duration $T=1000$. (a) $c=2$, symmetry  breaking and (b) $c=1$, symmetry conserving.

\item Fig. 5  Residence time  distribution $P(T_r)$ in the presence of weak  periodic signal with
$A=10^{-5}$ and $T=1000$. 
 Plot 1: $c=2$, symmetry breaking; plot 2: $c=1$,
symmetry conserving. 

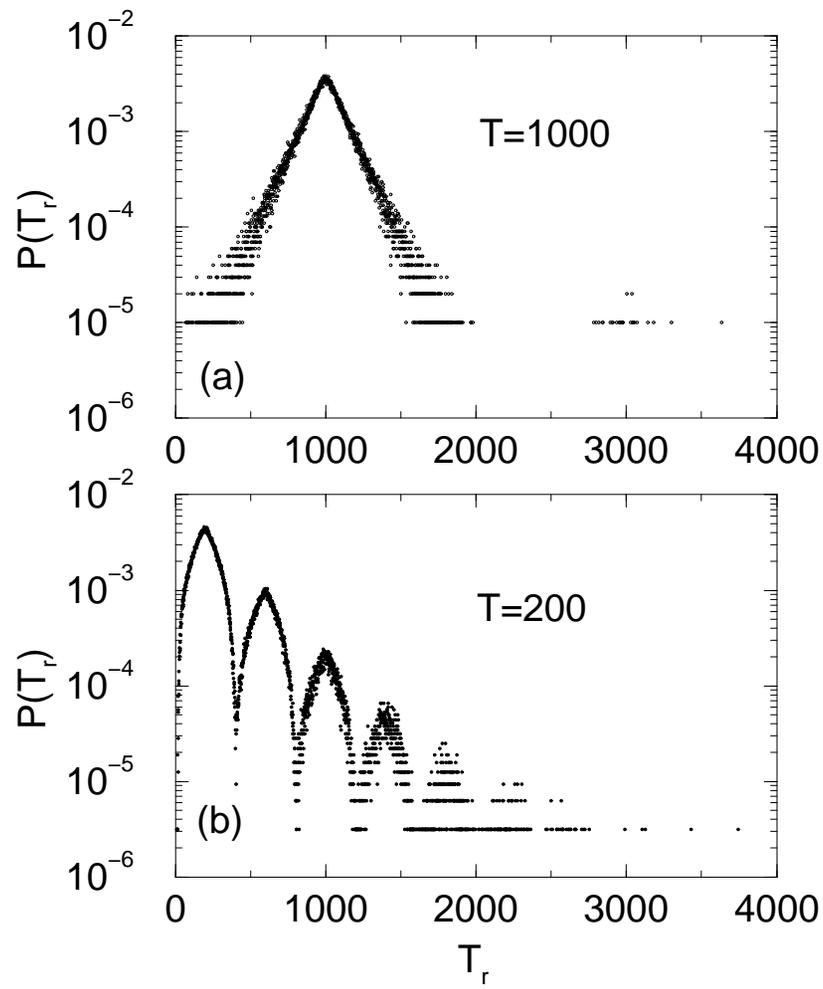
\item Fig. 6. Residence time distribution  $P(T_r)$ in the presence of weak  periodic signal with $A=10^{-5}$ 
but different bit duration $T$.

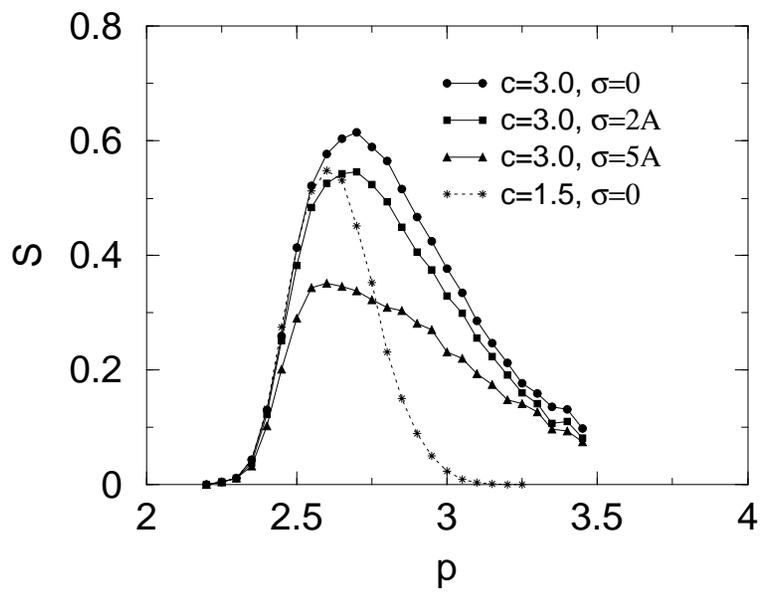
\item Fig. 7. The strength $S$ of the first peak as a function of $p$ for different level of additive noise. 
The input signal is $A=10^{-7}, T=1000$.

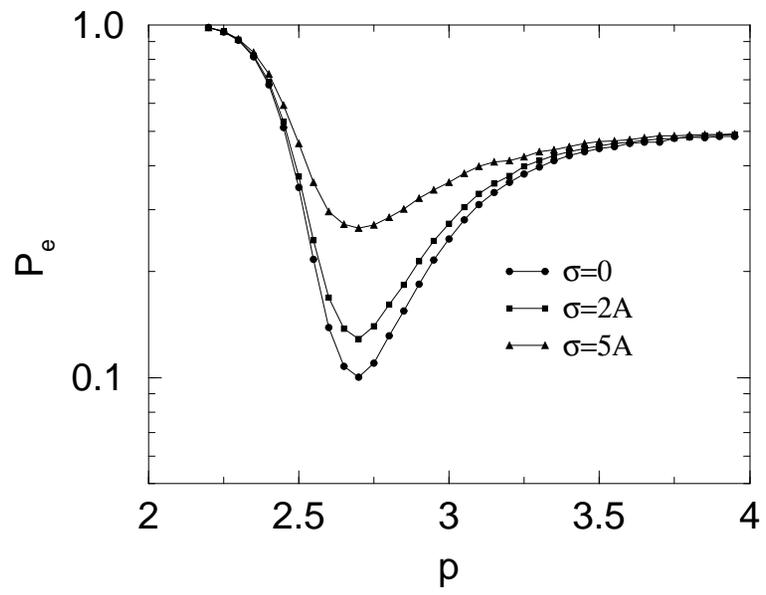
\item Fig. 8. Bit error probability $P_e$ as a function of $p$. The parameters are $c=3.0$, $A=10^{-7}, T=1000$.

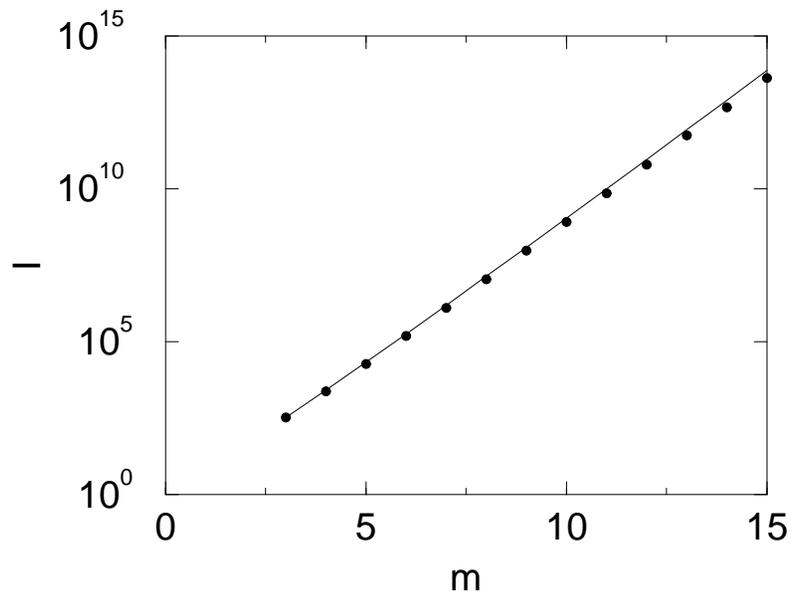
\item Fig. 9.   Amplification of the weak  signal with  $A=10^{-m}$ under the  adiabatic condition $T\gg T_0$.
               The amplification  factor (dots) is obtained by the ensemble average estimated  
               with $10^{4}$ samples of the time series of the system at $p=2.72$. The solid line is the  analytical  
               result of  Eq. (20)  with $y_{max}=p$.
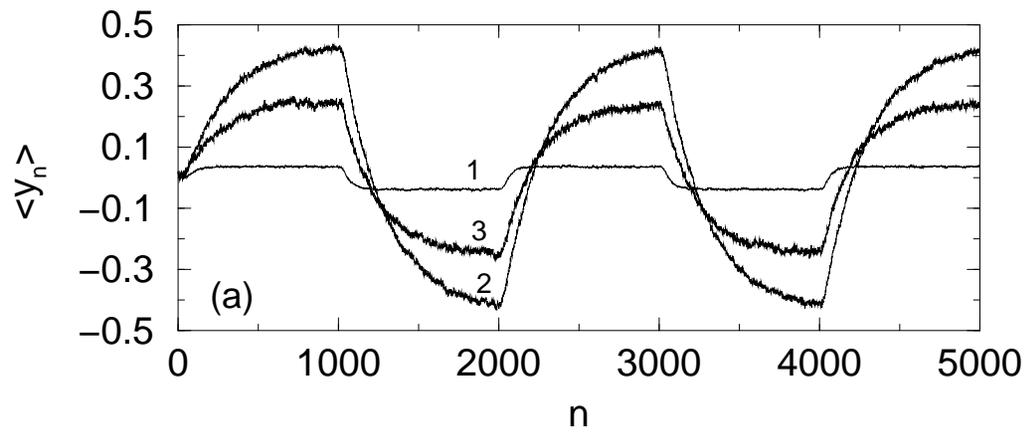
\item Fig. 10. Amplification property  as a function of the parameter $p$. 
              (a) Time series  of the ensemble average $\langle y_n \rangle $ for  several typical values of $p$. 
                 (1) $p=2.4$, (2) $p=3.1$, and (3) $p=3.4$. Other parameters are   
                  $A=10^{-5}$, $T=1000$ and $c=2.2$. 
              (b) $I$ as a function of $p$ for different values of $A$ and $c$. $T=1000$.

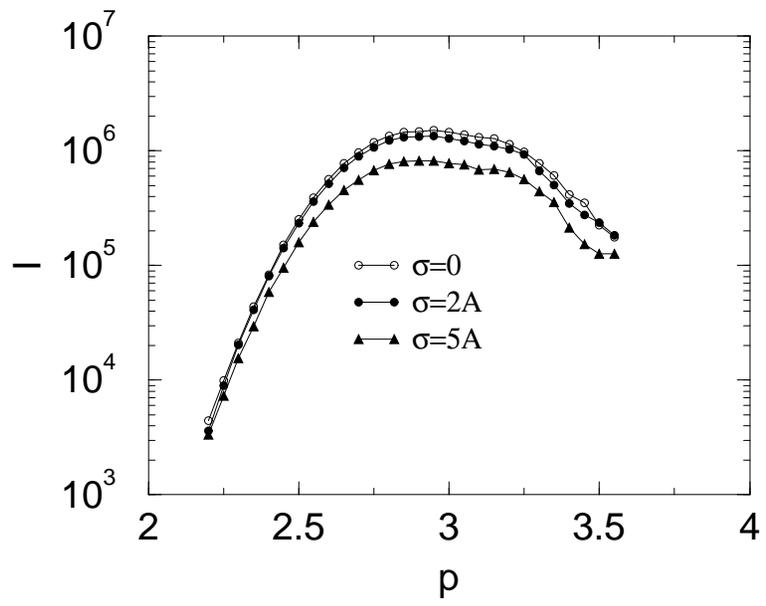
\item Fig. 11. Robustness of the  stochastic resonance to additive noise. 
               $A=10^{-7}$, $T=1000.$

\end{description}

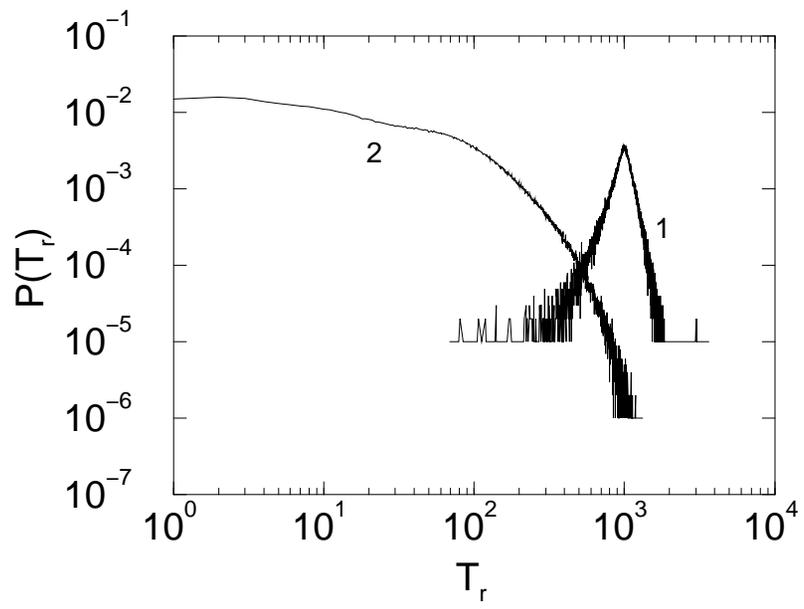
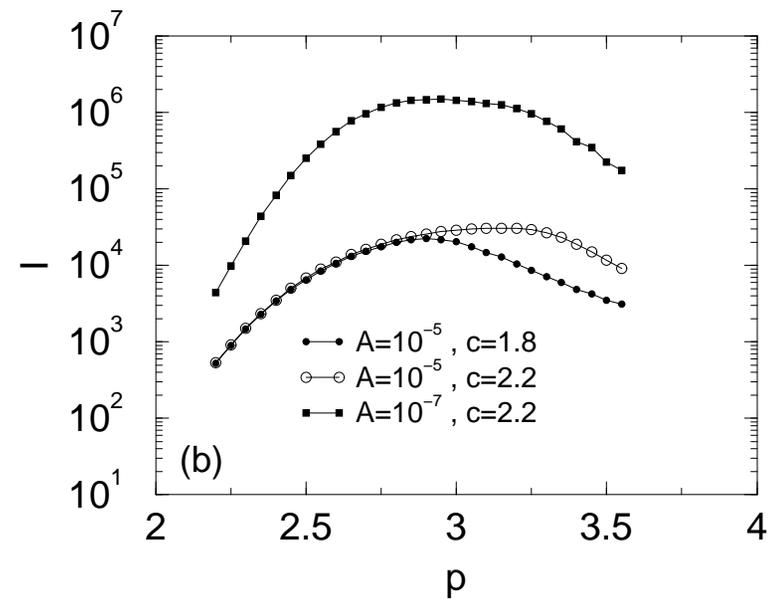
\end{document}